\documentclass[12pt,a4paper]{article}
\usepackage{amsfonts,amssymb}
\usepackage{amsmath}
\newcommand{\EQ}{\begin{equation}}
\newcommand{\EN}{\end{equation}}

\newtheorem{theo}{Theorem}[section]

\newtheorem{lem}{Lemma}
\newtheorem{prop}{Proposition}

\newtheorem{deff}{Definition}
\newtheorem{rem}{Remark}

\newcommand{\zero}{{\mathbf{0}}}

\newcommand{\wt}{\mbox{\rm wt}}
\newcommand{\Tr}{\mbox{\rm Tr}}

\newcommand{\bc}{{\bf c}}

\newcommand{\br}{{\bf r}}

\newcommand{\by}{{\bf y}}
\newcommand{\bx}{{\bf x}}

\newcommand{\bv}{{\bf v}}

\newcommand{\F}{\mathbb{F}}
\newcommand{\PG}{\mathbb{PG}}

\newcommand{\IA}{\operatorname{IA}}

\newcommand{\pr}{\indent{\bf Proof. \ }}
\newcommand{\qed}{\hspace*{5 mm}$\Box$}

\title{On the classification of completely regular codes with covering radius two
and antipodal dual}

\author{J. Borges, D. V. Zinoviev and V. A. Zinoviev}

\begin{document}
\maketitle

\centerline{\scshape Joaquim Borges}
\medskip
{\footnotesize
 \centerline{Department of Information and Communications Engineering}
 \centerline{Universitat Aut\`{o}noma de Barcelona}
} 

\medskip

\centerline{\scshape Dmitri Zinoviev}
\medskip
{\footnotesize
\centerline{A.A. Kharkevich Institute for Problems of Information Transmission}
 \centerline{Russian Academy of Sciences}
} 

\centerline{\scshape Victor Zinoviev}
\medskip
{\footnotesize
 \centerline{A.A. Kharkevich Institute for Problems of Information Transmission}
 \centerline{Russian Academy of Sciences}
} 

\bigskip

\begin{abstract}
We classify all linear completely regular codes which have
covering radius $\rho = 2$ and whose dual are antipodal.
For this, we firstly show several properties of such dual codes, which are two-weight codes.
\end{abstract}

\section{Introduction}
Let $\F_q$ be the finite field of order $q$, where $q$ is a prime power.
A $q$-ary $(n,N,d)_q$-code $C$ is a subset of $\F_q^n$ of size $N$ and {\em minimum (Hamming) distance} $d$. The {\em length} of $C$ is $n$.
A $q$-ary linear
$[n,k,d]_q$-code $C$ is a $k$-dimensional subspace of $\F_q^n$,
where $n$ is the length, $d$ is the minimum distance, and $q^k$ is the size of $C$.

The {\em packing radius} of a code $C$ is
 $e=\lfloor (d-1)/2 \rfloor$. Given any vector $\bv \in \F_q^n$, its
{\em distance to the code $C$} is $d(\bv,C)=\min_{\bx \in C}\{
d(\bv, \bx)\}$, where $d(\bv, \bx)$ denotes the distance between $\bv$ and $\bx$, and the covering radius of the code $C$ is
$\rho=\max_{\bv \in \F_q^n} \{d(\bv, C)\}$. Note that $e\leq \rho$.

Here we consider only {\em nontrivial}
linear codes, in particular, $[n,k,d]_q$-codes of dimension $2 \leq k \leq n-2$
and minimum distance $3 \leq d \leq n-1$.

For a given code $C$ of length $n$ and covering radius $\rho$,
define
\[
C(i)~=~\{\bx \in \F_q^n:\;d(\bx,C)=i\},\;\;i=0,1,\ldots,\rho.
\]
The sets $C(0)=C,C(1),\ldots,C(\rho)$ are called the {\em subconstituents} of $C$.

Say that two vectors $\bx$ and $\by$ are {\em neighbors} if
$d(\bx,\by)=1$. Denote by $\zero$ the all-zero vector.

\begin{deff}[\cite{N81}]\label{de:1.1} A code $C$ of length
$n$ and covering radius $\rho$ is {\em completely regular} (shortly
CR), if for all $l\geq 0$ every vector $x \in C(l)$ has the same
number $c_l$ of neighbors in $C(l-1)$ and the same number $b_l$ of
neighbors in $C(l+1)$. Define $a_l = (q-1){\cdot}n-b_l-c_l$ and
set $c_0=b_\rho=0$. The parameters $a_l$, $b_l$ and
$c_l$ ($0\leq l\leq \rho$) are called {\em intersection numbers}
and the sequence $\{b_0, \ldots, b_{\rho-1}; c_1,\ldots, c_{\rho}\}$
is called the {\em intersection array} (shortly $\IA$) of $C$.
\end{deff}

A more general class of codes is the class of uniformly packed codes in the wide sense:

\begin{deff}[\cite{bas1}]
Let $C$ be a binary code of  length $n$ and let $\rho$ be its
covering radius. We say that $C$ is {\em uniformly packed} in the
wide sense, i.e., in the sense of \cite{bas1}, if there exist
rational numbers $~\beta_0,\ldots,\beta_{\rho}~$ such that for any
$\bv\in\F^n$
\EQ\label{eq:2.1}
\sum_{k=0}^{\rho}\beta_k\,\alpha_k(\bv)~=~1,
\EN
where $\alpha_k(\bv)$ is the number of codewords at distance $k$ from
$\bv$.
\end{deff}

Note that the case $\rho=e+1$ and $\beta_{\rho-1} = \beta_{\rho}$
corresponds to {\em strongly uniformly packed} codes \cite{sem2},
the case $\rho=e+1$ corresponds to {\em quasi-perfect uniformly packed} codes
\cite{goet}, and the case $\rho=e$ and $\beta_i=1$ for all $i=0,1,\ldots, e$, corresponds to
{\em perfect} codes.


%

Completely regular codes are classical
subjects in algebraic coding theory, which are closely connected
with graph theory, combinatorial designs and algebraic
combinatorics. Existence, construction and enumeration of all such
codes are open hard problems (see \cite{BRZ19,BCN89,DKT,KKM,N81} and
references there).

All linear completely regular codes with covering radius $\rho=1$
are known \cite{B84,BRZ18}. The next case, i.e. completely regular
codes with $\rho = 2$, seems now too hard. Here we classify a
special class of such codes, namely, those linear completely regular
codes, whose dual codes are antipodal, i.e. their dual are
$[n,k,d]_q$-codes
with the following property: {\em for any two codewords $\bx$ and $\by$
the distance $d(\bx,\by)$ is one of two values $d$ and $n$}. In general, a $(n,N,\{d_1,d_2\})_q$-code $C$ is a code of length $n$, size $N$ and the distance between any two codewords $\bx$ and $\by$ verifies $d(\bx,\by)\in\{d_1,d_2\}$. Since we always consider distance-invariant codes, such a code has exactly $d_1$ and $d_2$ as weights and it is called a {\em two-weight} code. If $C$ is linear,
then we denote $C$ as a $[n,k,\{d_1,d_2\}]_q$-code (where $k$ is the dimension of $C$).

Our classification is based on the characterization of additive $(n,N,\{d,n\})_q$-codes (i.e. additive subgroups of $\F_q^n$) given in \cite{BDZZ22}.

The paper is organized as follows. In Section \ref{preliminary} we give some basic definitions and properties. In Section \ref{necessary}, we study some combinatorial properties of two-weight codes obtaining necessary conditions for the existence of such codes. In Section \ref{known}, we enumerate known families of linear completely regular codes with covering radius 2 and antipodal dual. Finally, in Section \ref{main}, we state the main theorem of the paper which establish that the enumeration in the previous section contains all linear completely regular codes with covering radius 2 and antipodal dual.


\section{Preliminary results}\label{preliminary}

In this section we see several results we will need in the next sections.

%
%

For a $[n,k,d]$-code $C$ let $(\eta^{\perp}_0, \ldots, \eta^{\perp}_n)$
be the weight distribution of its dual $[n,n-k,d^{\perp}]$-code
$C^{\perp}$. Assume $(\eta^{\perp}_0, \ldots, \eta^{\perp}_n)$ has
$s = s(C)$ nonzero components
$\eta^{\perp}_i$ for $1 \leq i \leq n$. Following to
Delsarte \cite{D73}, we call $s$ the {\em external distance} of $C$.

\begin{lem}\label{lem:2.1}
Let $C$ be a code with  covering radius
$\rho$ and external distance $s$. Then:
\begin{enumerate}
\item[(i)]\cite{D73} $\rho \leq s$.
\item[(ii)]\cite{bas2} $\rho = s$ if and only if $C$ is uniformly packed in the wide sense.
\item[(iii)]\cite{BCN89} If $C$ is completely regular, then it is uniformly packed in the wide sense.
\end{enumerate}
\end{lem}

As a consequence, if $C$ is a completely regular $[n,k,d]_q$-code, then the dual code $C^\perp$ is a two-weight $[n,n-k,\{d_1,d_2\}]_q$-code. Moreover, if $C^\perp$ is antipodal, then it is a $[n,n-k,\{d^\perp,n\}]_q$-code (where $d^\perp$ is the minimum distance of $C^\perp$). This is the class of completely regular codes that we classify in this paper.

\begin{deff}\label{deff:dm}
Let $G$ be an abelian group of order $q$ written additively. A square
matrix $D$ of order $q\mu$ with elements from $G$ is called a
difference matrix and denoted
$D(q, \mu)$, if the component-wise difference of any two different
rows of $D$ contains every element of $G$ exactly
$\mu$ times.
\end{deff}

Clearly the matrix $D$ is invariant under addition of the vector
$(a,a, \ldots,a)$, where $a \in G$, to any row of $D$. By doing such
operation we can obtain the {\em normalized} difference matrix
which has the zero first row and the zero first column.

From \cite{SZZ69} we have the following result.

\begin{lem}\label{lem:2.2}
For any prime power number $q$ and any natural numbers $\ell$ and $h$
there exists a difference matrix $D(q^\ell, q^h)$.
\end{lem}

We describe briefly the construction of all such difference matrices
$D(q^\ell, q^h)$ from \cite{SZZ69}. For any natural
numbers $\ell$ and $h$, denote $u = \ell + h$. For the Galois field
$\F_{q^u}$ with elements $\{f_0=0,f_1=1,f_2, \ldots, f_{q^u-1}\}$
denote by $F = [f_{i,j}]$
the matrix of size $q^u \times q^u$, whose
rows and columns are indexed by the elements of $\F_{q^u}$,
where $f_{i,j} = f_i f_j$, i.e. $F$ is a multiplicative table of the
elements of $\F_{q^u}$. For any natural $m$ fix a one-to-one correspondence
between elements of $\F_{q^m}$ and the vector space $\F_q^m$. Define
the operator $\Phi=\Phi_{u\rightarrow \ell}$, which map elements
$x = (x_1,\ldots,x_u)$ of $\F_q^u$ into elements
$x^{(\ell)} = (x_1,\ldots,x_\ell)$ of $\F_q^\ell$ by cutting the last
right $u-\ell$ positions of vectors from $\F_q^u$:
\[
\Phi(x_1,\ldots,x_\ell,\ldots, x_u) = (x_1,\ldots,x_\ell).
\]
Denote by $F^{[\ell]}$ the matrix obtained from $F$ by applying the
operator $\Phi$ to the all elements of $F$,
\[
F^{[\ell]} = [f^{[\ell]}_{i,j}]:\;f^{[\ell]}_{i,j} = \Phi_{u\rightarrow \ell}(f_{i,j}).
\]

\begin{lem}\label{lm:2.3}
For any prime power number $q$ and any natural numbers $\ell$ and $h$
the matrix $F^{[\ell]}$ is an additive difference matrix $D(q^\ell, q^h)$.
If $\ell$ divides $h$, i.e. $N = q^{h/\ell + 1}$ then $D$ is linear.
\end{lem}

Now, we explain the construction of $(n,N,\{d,n\})_q$-codes based on difference
matrices. Without loss of generality, we can write
$G=\{0,1, \ldots, q-1\}$. Assume that the first row of $D$
consists of zeros. Denote by $D^{(g)}$ the
matrix obtained from $D$ by adding the element $g \in G$ to
all elements of $D$, i.e. if $D = [d_{i,j}]$, then $D^{(g)} =
[d_{i,j} + g]$ for all $i$ and $j$ (the addition is in $G$).
By the definition of $D$ the matrix $D^{(g)}$ is a difference
matrix $D(q,\mu)$. It follows also that for any two rows $\br$
from $D$ and $\br^{(g)}$ from $D^{(g)}$ the following property is
valid \cite{SZZ69}:
\EQ\label{eq:3.1}
d(\br, \br^{(g)}) =
\left\{
\begin{array}{ccc}
q \mu,\;\;&\mbox{if}&\;\; \br^{(g)} = \br + (g,g, \cdots, g),\\
(q-1)\mu,\;\;&\mbox{if}&\;\; \br^{(g)} \neq \br + (g,g, \cdots, g).\\
\end{array}
\right.
\EN
Clearly the matrix $D(q, \mu)$ induces an equidistant $(q\mu-1,
q\mu, \mu(q-1))_q$-code, which is optimal with respect to the Plotkin
upper bound
\EQ\label{eq:2.20}
N \leq \frac{qd}{qd-(q-1)n},
\EN
if the denominator is positive.
To see it, first, we have to transform $D$ to the form, which has
the trivial column and, second, delete this trivial column.
From (\ref{eq:3.1}) we deduce

\begin{lem} $\cite{SZZ69}$
The rows of the $(N\times n)$-matrix
$[D^{(0)}\,|\,\cdots \,|\,D^{(q-1)}]^t$ form a
two-weight $(n,N,\{d,n\})_q$-code with parameters
\EQ\label{eq:2.21}
n = q\mu,\;\;N = q^2\mu,\;\;d = \mu(q-1).
\EN
\end{lem}
The code $C$ based on a difference matrix $D$ is called a {\em difference
matrix code}, or, shortly, DM-{\em code}.
Any $(n,N,\{d,n\})_q$-code whose parameters satisfies (\ref{eq:2.21})
is called a {\em pseudo difference matrix code}, or, shortly, PDM-{\em code}.
Later we will see that an additive PDM-code is a DM-code.
These codes are optimal with respect to the $q$-ary analog of Gray-Rankin bound
\cite{BDZH06}. Any $q$-ary $(n,N,\{d,n\})_q$-code, which can be partitioned into
trivial $(n,q,n)_q$-subcodes, satisfies
\EQ\label{eq:2.3}
\frac{N}{q} \leq \frac {q(qd - (q-2)n)(n-d)}{n-((q-1)n-qd)^{2}},
\EN
provided $n-((q-1)n-qd)^{2} > 0$.

We also recall the bound for the cardinality $N$ of the
code $C$ when the maximal distance between codewords is limited,
say by $D$ (see \cite{HKL06} for small cases and \cite{BDHSS-TIT}  
for the general case). For $D=n$ this bound looks as follows:
\EQ\label{eq:2.6}
N \leq \frac {q^{2}d}{dq-(q-1)(n-1)},
\EN
if the denominator is positive.

Recall that a $q$-ary matrix $M$ of size $N \times n$ is called an
orthogonal array denoted $OA(N,n,q,t)$ of strength $t$, index
$\lambda = N/q^t$ and $n$ constraints, if every $N \times t$ submatrix of it
contains as a row every $q$-ary vector of length $t$ exactly $\lambda$
times (see \cite{BJL86}).

%

\section{Necessary conditions}\label{necessary}

As we already mentioned a linear CR $[n,k,d]_q$-code $C$ with
covering radius $\rho=2$ has a dual code $C^\perp$ which is
a linear $[n,k^\perp=n-k,d^\perp]_q$-code with the following
property: for any codeword $\bc \in C^\perp$ its (Hamming)
weight has only two possible values, namely, $\wt(\bc) \in \{w_1, w_2\}$,
where $w_1 = d^\perp$. More generally, the number of nonzero weights of $C^\perp$ (called {\em external distance} of $C$) equals the covering radius of $C$, whenever $C$ is completely regular (see Lemma \ref{lem:2.1}).

Here we consider the case $w_2 = n$, i.e.
the code $C^\perp$ is antipodal.
So in order to classify linear CR codes with $\rho=2$ whose dual are
antipodal, we have to classify all codes which have
only two weights $d$ and $n$, i.e. all $[n,k,\{d,n\}]_q$-codes. Such classification is given in \cite{BDZZ22}.
The natural question for existence of a $q$-ary  two-weight
$(n,N,\{d, n\})_q$-code is under which conditions such code can exist. In \cite{BDZZ22} this question is answered and
here we study the particular case for linear such codes.

Let $\PG(n,q)$ denote the $n$-dimensional projective space over
the field $\F_q$. A {\em $m$-arc} of points in $\PG(n,q)$,\, $m
\geq n+1$ and $n \geq 2$, is a set $M$ of $m$ points, such that no
$n+1$ points of $M$ belong to a hyperplane of $\PG(n,q)$. The
$(q+1)$-arcs of $\PG(2,q)$ are called {\em ovals}, the $(q+2)$-arcs
of $\PG(2,q)$,\,$q$ even, are called {\em complete ovals}  or {\em
hyperovals} (see, for example, \cite{D69,T95}).

A linear code $C$ is called {\em projective} if its dual code
$C^\perp$ has minimum distance $d^\perp \geq 3$ (i.e., any
generator matrix of $C$ does not contain two columns that are
scalar multiples of each other). 

Let $C$ be a projective $[n, k, d]_q$-code with nonzero weights $w_1, w_2,\ldots, w_s$ and generator matrix $G$. For $\alpha$ and $\beta$
such that $\alpha w_i + \beta$ are nonnegative integers for all
$i$, we can define a dual transform, say $C^*$, of C in the
following way. Consider all nonzero vectors
$v \in \F^k_q$ for which the corresponding points in $\PG(k-1,q)$
are different. A matrix $G^*$ is constructed so that it
contains as column vectors all such vectors $v$ taken
$w(vG) \times \alpha + \beta$ times. This matrix $G^*$ is the generator
matrix of a two-weight code $C^*_{\alpha,\beta}$ which we call the
{\em projective dual} of $C$. Therefore, a two-weight code has always a projective dual code.

For projective  $[n,k,d]_q$-codes $C$
one can also define the concept of {\em complementary code} (see, for
example, \cite{CK86}).

Let $[C]$ denote the matrix formed by all codewords of $C$ (i.e.,
the rows of $[C]$ are codewords of $C$).
The code $C_c$ is called a complementary of $C$, if the matrix
$[[C]\,|\,[C_c]]$ is a linear equidistant code and $C_c$ has the
minimal possible length which gives this property.

We recall a result in \cite{D71}. Denote $n_m = (q^m-1)/(q-1)$.
For a given $[n,k,d]_q$-code
$C$ with parity check matrix $H$ define its {\em complementary}
$[n_{n-k} -n,k,\bar{d}]$-code $\bar{C}$, whose parity check matrix
$\bar{H}$ is obtained from the matrix $H_{n-k}$ (the parity-check matrix of a Hamming code of length $n_{n-k}$) by removing all
the columns of $H$ and multiples of them. Recall an important
property of complementary codes: {\em to any codeword of weight
$w$ in a $[n,k,d]_q$-code $C$ corresponds a codeword of weight
$\bar{w}=q^{n-k-1} - w$ in the complementary code $\bar{C}$}. As a
corollary of this fact above we have the next lemma.

\begin{lem}\label{complementary1} $\cite{D71}$
A linear $[n,k,d]_q$-code $C$ with covering radius
$\rho = 2$, which is not the dual of a difference matrix code,
does exist simultaneously with its complementary projective code
$\bar{C}$ with the same covering radius $\bar{\rho} = 2$.
\end{lem}

The extension of this well-known concept to arbitrary linear
two-weight $[n,k,\{d,d+\delta\}]_q$-codes was given in \cite{BDZZ21}.
Here we give a variant of such lemma for the case of
$[n,k,\{d,n\}]_q$-codes. Somehow, it is connected with {\em anticodes}
due to Farrel \cite{F70} and {\em minihypers} (see \cite{HH01}).

\begin{lem}\label{complementary2} $\cite{BDZZ21}$
Let $C$ be a $q$-ary linear nontrivial two-weight
$[n,k,\{d,n\}]_q$-code, which is not the dual of a
$s$-times repetition of a difference matrix code,
and let $\mu_1$ and $\mu_2$ denote the
number of codewords of weight $d$ and $n$, respectively. Then
there exist the complementary linear two-weight $[n_c,k,\{d_c,
d_c + \delta\}]_q$-code $C_c$, where
\[
n + n_c = s\,\frac{q^k-1}{q-1},\;\;d + d_c + \delta = s q^{k-1},\;\;
n = d + \delta,\;\;s=1,2,
\ldots ,
\]
and where $C_c$ contains $\mu_1$
codewords of weight $d_c + \delta$ and
$\mu_2$ codewords of weight $d_c$ and where $C_c$ is of minimal
possible length $n_c$, such that the matrix $[[C]\,|\,[C_c]]$ is an
equidistant $[s(q^k-1)/(q-1), k, sq^{k-1}]_q$-code.
\end{lem}

Note that the integer $s$ in Lemma \ref{complementary2} is the maximal
size of the collection of columns in the generator matrix of $C$
which are scalar multiples of one column. For projective
two-weight $[n,k,\{w,n\}]_q$-codes (i.e. for the case $s=1$) the
following results are known.

\begin{lem}\label{lm:4.0} $\cite{D72}$
Let $C$ be a two-weight projective $[n,k,\{w, n\}]_q$-code over
$\F_q$,\,$q = p^m$,\, $p$ is prime. Then there exist two integers
$u \geq 0$ and $h \geq 1$, such that
\[ w = h\,p^u,\;\;n = (h+1)\,p^u. \]
\end{lem}

For the projective case, we recall the following result (which
directly follows from the MacWillams identities, taking into
account that the dual code $C^\perp$ has minimum distance $d^\perp
\geq 3$) (see \cite{D72}).

\begin{lem}\label{lm:4.2}
Let $C$ be a two-weight projective $[n,k,\{w, n\}]_q$-code $C$
over $\F_q$,\,$q = p^m$,\, $p$ is prime. Denote by $\mu_1$ the
number of codewords of $C$ of weight $w$ and by $\mu_2$ the number
of codewords of weight $n$. Then

\EQ \label{lm4-eq1}
\left.
\begin{array}{llc}
 w\,\mu_1 + n\,\mu_2 &=& n(q-1)q^{k-1},\\
w^2\,\mu_1  + n^2\,\mu_2 &=& n(q-1)(n(q-1)+1)q^{k-2}.
\end{array}
\right\}
\EN
\end{lem}

In \cite{BDZZ21} (see also \cite{BDZZ20} for the special case
$n-d = 1$)
integrality conditions are derived, similar to conditions obtained
by Delsarte in \cite{D72} (see also \cite{CK86})
for projective two-weight codes using simple
combinatorial arguments not connected with eigenvalues of strongly
regular graphs. For the case of arbitrary two-weight
$(n,N,\{d, n\})_q$-codes with distances
$d$ and $n$ that conditions reduces to the following result which can be found in \cite{BDZZ22}.

\begin{theo}\label{th:4.1}
Let $C$ be a
non-trivial $q$-ary two-weight $(n,k,\{d, n\})_q$-code. Then
\begin{itemize}
\item [(i)] The code $C$ has cardinality $N=q^k$ such that
\EQ\label{eq:3.22}
\max\{(q-1)n + 1,\;q^2\} \;\leq\; N \;\leq \;\frac{q^2d}{qd-(q-1)(n-1)}\,.
\EN

\item [(ii)] The right inequality in (\ref{eq:3.22}) is an
equality if and only if the matrix $[C]$, formed by the all
codewords of $C$, is an orthogonal array of strength $t \geq 2$.

\item [(iii)] If the right inequality in  (\ref{eq:3.22}) is
an equality, the length $n$ and the
distance $d$ of the code $C$ look as follows:
\EQ\label{eq:3.23}
n = \frac{N(q(d+1) - 1) - q^2d}{N(q-1)}\,.
\EN
and
\EQ\label{eq:3.24}
d = (n-1) \cdot \frac{(q-1)N}{q(N-q)}\,.
\EN

\item [(iv)] The left inequality in (\ref{eq:3.22}) is an equality,
if and only if, either $C$ is a Latin-square code with parameters
$(n,q^2,\{n-1,n\})_q$, where $n \leq q$, or $C$ is an equidistant
$(n,N,d)_q$-code (i.e. the case which we avoid).

\item [(v)] If $C$ is a nontrivial two-weight $(n,q^2,\{d,n\})_q$-code,
then the number $N$ divides $q^2d$ and the number $(q-1)$
divides $(N-1)d$.
\end{itemize}
\end{theo}

In the next statement we formulate the variant of Theorem 2 from \cite{BDZZ21}
for the case of nontrivial $[n,k,\{d,n\}]_q$-codes (hence what does not
need a proof).  Here we assume that $q = p^m$ where $m \geq 1$ and
$p$ is prime. Given $q=p^m$ and an arbitrary positive integer $a$ denote
by $\gamma_a \geq 0$ the maximal integer, such that $p^{\gamma_a}$ divides $a$,
i.e. $a = p^{\gamma_a}\,h$, where $h$ and $p$ are co-prime.
Let $\gamma_d$, $\gamma_\delta$ and $\gamma_c$ be defined similarly for $d$,
$\delta$ and $d_c$, respectively. Recall that $(a,b)$ denote the greatest
common divisor (gcd) of two integers $a$ and $b$.

\begin{theo}\label{th:4.2}
Let $q = p^m$, where $m \geq 1$ and $p$ prime. Let $C$ be a $q$-ary linear
(two-weight) $[n,k,\{d, n\}]_q$-code of dimension $k \geq 2$ and let $C_c$ be its
complementary two-weight $[n_c,k,\{d_c, d_c+\delta\}]_q$-code $C_c$, where
\[
d + \delta = n\;\;\mbox{and}\;\;d + d_c + \delta = s\,q^{k-1},\;\;s \geq 1\,.
\]
\begin{itemize}
\item[(i)] If $s = 1$ and $k \geq 4$, i.e. $C$ and hence $C_c$ are projective
codes, then the following two equalities  are satisfied:
\EQ\label{eq:4.0}
(q,d) \,=\,(q,\delta)\;\;\mbox{and}\;\;(q,d_c) \,=\,(q,\delta)\,.
\EN

\item[(ii)] If $s = 1$ and $k = 3$, then both equalities in (\ref{eq:4.0}) are
satisfied, if at least one  of the following two conditions takes place:
\[
(d,q)^2 \leq q (n(n-1),q)\;\;\mbox{or}\;\;(d+\delta,q)^2 > q (n_c(n_c-1),q)\,.
\]
\item[(iii)] If $s = 1$ and $k \geq 2$, then at least one of the following two
equalities is satisfied:
\EQ\label{eq:4.00}
\gamma_d = \gamma_\delta,\;\;\mbox{or}\;\;\gamma_c = \gamma_\delta\,.
\EN
\item[(iv)] If $s \geq 1$ and $k \geq 3$, then at least one of the two equalities
in (\ref{eq:4.00}) (respectively, in (\ref{eq:4.0})) is valid.

\end{itemize}
\end{theo}

\section{Known completely regular codes with covering radius $\rho=2$}\label{known}

Now, we enumerate all linear nontrivial $[n,k,\{d,n\}]_q$-codes,
which give the classification of linear CR codes with $\rho=2$ and antipodal dual.
Most of these two-weight codes can be found in
the comprehensive survey of two-weight codes by Calderbank and Kantor
\cite{CK86}) and all such codes are given in \cite{BRZ10}.

First we give one statement, which is
a reformulation of the corresponding result from \cite{BRZ10}.

\begin{theo}\label{theo:4.1}
Let we have a nontrivial linear $[n,k,d]_q$ code $C$. Let $G$
be its generating matrix. Then, $C$ is an $[n,k,\{d,n\}]_q$-code,
if and only if the matrix $G$ looks, up to equivalence, as follows:
\[
G=\left[
\begin{array}{cccc}
\,1 \,&\,\cdots \,&\,1\,\\
\, &\,\,G^*\,\,&\,\,\\
\end{array}
\right],
\]
where $G^*$ generates an equidistant code $C^*$ with the following
property: for every nonzero codeword $\bv\in C^*$, every symbol
$\alpha\in\F_q$, which occurs in a coordinate position of $\bv$,
occurs in this codeword exactly $n-d$ times, where
$d$ is the minimum distance of $C^*$.
\end{theo}

Following to \cite{BDZH06} call a trivial $(n,q,n)_q$-code by a
{\em simplex}. Say that a $q$-ary distance invariant code of
length $n$ is a {\em simplex code} if it contains as a subcode
a simplex, i.e. a $(n,q,n)_q$-code. Clearly an additive
$(n,N,\{d,n\})_q$-code is a distance invariant simplex code.
The following result can be found in \cite{BDZH06}.

\begin{prop}\label{simplex}
Let the cardinality of a $q$-ary code $C$ of length $n$ and with
distance $d=\frac{(q-1)n}{q}$ be equal to $qn$. Then the code
$C$ can be presented as a union of non-intersecting simplex codes.
\end{prop}
\bigskip

The natural question is {\em under which conditions a simplex
code from Proposition \ref{simplex} is a PDM, or DM code}. A
partial answer is given by the following

\begin{theo}[\cite{BDZZ22}]\label{theo:4.2}
Let $C$ be a distance invariant simplex $(n,N,\{d,n\})_q$-code.
Then
\begin{enumerate}

\item[(i)] The code $C$ can be partitioned into non-intersecting
subcodes as follows:
\[
C = \bigcup_{i=1}^{N/q} C_i,
\]
where $C_i$ for every $i$ is a simplex and $N$ is a multiple of $q$.

\item[(ii)] For any codeword $\bc \in C$, which is not of the
type $(a,a, \ldots, a)$, \,$a \in \F_q$, every symbol
$\alpha \in \F_q$, which occurs in a coordinate position of $\bc$,
occurs in this codeword exactly $\mu$ times, where $\mu = n-d$,
and $n$ is a multiple of the number $\mu$.

\item[(iii)] The distance $d$ of $C$ satisfies the following
inequality:
\EQ\label{dbound}
d \leq n \cdot \frac{q-1}{q}.
\EN

\item[(iv)] If (\ref{dbound}) is an equality, and $N = q n$, then
the code $C$ is a pseudo difference matrix code with parameters
\[
n = \mu q,\;\;,N = \mu q^2,\;\;d = \mu (q-1),\;\;\mu = n-d.
\]

\item[(v)] If in (iv) $C$ is additive then $C$ is a difference matrix
code.

\end{enumerate}
\end{theo}

\begin{rem}\label{rem:1}
The conditions $n=q(n-d)$ and $N=q n$ can not be removed in (iv)
and (v). Consider the matrix $[C] = [D^{(0)}\,|\,\cdots \,|\,D^{(q-1)}]^t$
formed by translates $D^{(i)}$ of a difference matrix $D=D(q,\mu)$,
where $C$ is a DM $(n,N,\{d,n\})_q$-code. If we
remove any one (or more) such matrices $D^{(i)}$ from the matrix $[C]$
of codewords we obtain a distance invariant simplex code with some
cardinality $N^* < qn$, i.e. a nonlinear two-weight
$(n,N^*,\{d,n\})_q$-code, which satisfies the conditions of the theorem.
Similarly, we can not remove the
condition $N = qn$ in (iv) and (v). For example, the linear Bose-Bush codes
(see below) have length $n < q(n-d)$.  Similarly,
an additive $(n,N,\{d,n\})_q$-code should not necessary be of
cardinality $q^k$. For example, the difference matrix $D(4,2)$
induces the optimal additive $(8,32,\{6,8\})_4$-code with cardinality
$N \neq 4^k$.
\end{rem}

\begin{rem}\label{rem:2}
The case of codes with $N = q^2$ is quite specific. A well known result
is that $r-2$ mutually orthogonal Latin squares of order $q$ induces
a $(r,q^2,\{r-1,r\})_q$-code. For the case when $q$ is a prime power
there exist $q-1$ mutually orthogonal Latin squares which provides
a linear equidistant $[q+1,2,q]_q$-code (the inverse statement is also
valid for any length $r$ and well known). Using these codes with
corresponding $r$, we can
build, in an evident way (using partitions into simplex codes), a
$(n,2,\{d,n\})_q$-code for any natural $d$ and $n$ with the only
condition, that $d \geq n/2$, which
is guarantied by the Plotkin bound (\ref{eq:2.20}). So, we avoid
all these trivial codes.
\end{rem}

Now we give the known families of linear nontrivial CR
$[n,k,d]_q$-codes which were all mentioned in \cite{BRZ10}:

\begin{enumerate}
\item[{\bf $(CR.1)$}] {\em Binary extended perfect Hamming
codes}. These  are
$[n=2^m,k=n-m-1,4]$-codes, the most famous binary codes of
difference matrix type, dual to linear Hadamard
$[2^m, m+1, 2^{m-1}]$-codes.
These codes have parity check matrix $H_h$ of size
$(m+1) \times 2^m$, m rows are formed by all $2^m$ different binary column
vectors of length $m$ and the $(m+1)$th row is the
all-ones vector ${\bf 1}=(1, 1, 1, \ldots, 1)$.
For this case, the dual to Hadamard codes  are
extended binary perfect Hamming codes.

These codes have IA: $\{n,n-1;1,n\}$.

\item[{\bf $(CR.2)$}] {\em Dual to the Difference matrix codes.}
These are $[n=q^m, n-m-1, 3]_q$-codes  \cite{SZZ69} and
induced by difference matrices. The wide class of such matrices and
corresponding additive and linear DM codes with
parameters (\ref{eq:2.21}) was given in \cite{SZZ69} (see Lemma \ref{lem:2.2}),
for the values
\[
q = p^{mh},\;\;\mu = p^{m\ell},\;\mbox{where\;$p$ is prime and $m = 1,2,3,\ldots $}
\]
for any natural numbers $h$ and $\ell$.
Lemma \ref{lem:2.2} provides a simple construction of all
these codes and Theorem \ref{theo:4.2} gives a description of such
codes.

These codes have IA: $\{n(q-1),n-1;1,n(q-1)\}$.

\item[{\bf $(CR.3)$}] {\em Dual to Latin squares codes, which are also
MDS-codes.} These are common known $[n,n-2,3]_q$-codes, induced by
$n-2$ mutually orthogonal Latin squares of order $q$. In Theorem \ref{th:4.1}
the Latin squares $[n,2,\{n-1,n\}]_q$-codes correspond to the distance
$d = n-1$, which gives $N=q^2$.
Since $q$ is a prime power, then a linear Latin square $[n,2,\{n-1,n\}]_q$-code $C_\ell$ has any length $n$ in the region $2 \leq n \leq q$ and it is
generated by the matrix $G_\ell$,
\[
G_\ell = \left[
\begin{array}{ccccc}
&1,\,1,\; 1,\;  1,\;    &\ldots,& \;1&\\
&0,\,1,\,a_2,\,a_3,\,   &\ldots,& a_{n-1}&
\end{array}
\right],
\]
where $a_0=0, a_1=1, a_2, \ldots, a_{n-1}$ are different elements of
$\F_q$.

These codes have IA: $\{n(q-1),(q-n+1)(n-1);1,n(n-1)\}$.

\item[{\bf ${(CR.4)}$}] {\em Dual to Bose-Bush codes, or extended Hamming
codes.} These are $[q+2,q-1,4]_q$-code, whose dual  were constructed
by Bush \cite{B52} in 1952 and which exist for any $q = 2^m \geq 4$
(the family $TF1$ in \cite{CK86}). In Theorem \ref{th:4.1} the Bose-Bush codes
correspond to the distance $d = n-2$, which gives $N=q^3$. This is a
classical object in projective geometry, since such a code is induced by
a hyperoval in $PG(3,q)$. The dual $[q+2,q-1,4]_q$-codes are extended
$q$-ary (perfect) Hamming codes. We give the generator matrix $G_b$
for the  Bose-Bush $[q+2,3,\{q,q+2\}]_q$-code which differs from the matrices,
given by Bush \cite{B52} and Delsarte \cite{D72}, since we want to follow
the form given by Theorem \ref{theo:4.1}, namely,
\EQ\label{bushcode}
G_b = \left[
\begin{array}{ccccccccl}
1\;&1\;&1\;&1\;&1\,      &\ldots &\,1\,  &\ldots  &\;1\\
0\;&1\;&0\;&1\;&x_1\,    &\ldots &\,x_i\,&\ldots  &\;x_{q-2}\\
0\;&0\;&1\;&1\;&y_1\,    &\ldots &\,y_i\,&\ldots  &\;y_{q-2}
\end{array}
\right],
\EN
where
\[
x_i=\frac{\alpha^i}{1+\alpha^i+\alpha^{2i}},\;\;
y_i=\frac{\alpha^{2i}}{1+\alpha^i+\alpha^{2i}}\,.
\]

These codes have IA: $\{(q+2)(q-1),q^2-1;1,q+2\}$.

\item[{\bf $(CR.5)$}] {\em Dual to Delsarte codes,} which are
$[n = q(q-1)/2, k = n-3, 4]_q$-codes. For the case
$q=2^s \geq 4$ Delsarte in \cite{D72} constructed the $q$-ary
$[q(q-1)/2, 3, \{q(q-2)/2, q(q-1)/2\}]_q$-codes.
These codes are projective dual to the codes $(CR.4)$.
(see family $TF1^d$ in \cite{CK86}). In Theorem \ref{th:4.2} this case
corresponds to the distance $d = n(q-2)/(q-1)$, which implies $N = q^3$.
The dual codes are extended $q$-ary CR codes, which can be generated
by the matrix $G_d$, which is a submatrix of $G_b$, formed by the columns
$[1,\,\alpha^i,\,\alpha^{2i}]^t$, such that $\Tr(\alpha^{3i}) = 1$, where
$\Tr(x)$ is the trace from $\F_q$ to $\F_2$, i.e.
\[
\Tr(x) = x + x^2 + x^4 + \cdots + x^{q/2}\,.
\]

These codes have IA: $\{(q-1)n,(q-2)(q+1)(q+2)/4;1,q(q-1)(q-2)/4\}$.

\item[{\bf $(CR.6)$}] {\em Dual to Denniston codes.} These are
$[n = 1 + (q+1)(h-1), k = (q+1)(h-1) - 2, d = 4]_q$-codes, which are
dual to Denniston $[n = 1 + (q+1)(h-1), 3,\{q(h-1), n\}]_q$-codes, where
$1 < h < q$ and $h$ divides $q$, for $q=2^r \geq 4$ (the family
$TF2$ in \cite{CK86}). In Theorem \ref{th:4.1} this case
corresponds to the distance $d = (n-1)q/(q+1) = n-h+1$, which
implies $N = q^3$. The dual codes are extended $q$-ary
codes with $d=4$. For
$h=2$ we obtain the Bose-Bush codes $(CR.4)$. The codes $(CR.6)$ are
formed by maximal arcs in projective plane \cite{D69}. The
construction corresponds to Theorem \ref{theo:4.1}. We explain
shortly how to construct such a code for arbitrary $q = 2^m \geq 4$
and natural $h \geq 2$, which divides $q$, i.e. $h = 2^u \leq q/2$.
For a given $\F_q$ let $H$ be a subgroup of order $h$ of the
additive group of $\F_q$. Let $\varphi(x,y)$ be an irreducible
quadratic form over $\F_q$, say, $\varphi(x,y)=ax^2+bxy+cy^2$.
Then the Denniston $[n,3,\{d,n\}]_q$-code is generated by the
following $(3 \times n)$-matrix
\EQ\label{gen-denn-codes}
G_d = \left[
\begin{array}{cccc}
1   \;&1    & \;  \cdots & \;1\\
x_1 \;&x_2  & \;  \cdots & \; x_n\\
y_1 \;&y_2  & \;  \cdots & \; y_n\\
\end{array}
\right],
\EN

where $n = (q+1)((h-1)+1)$, \;$d = n-h$, and where $(x_i,y_i)$
are the all ordered pairs of elements of $\F_q$ such that
$\varphi(x_i,y_i) \in H$. If the sets of all elements $\{x_i\}$ and
$\{y_i\}$ do not suit to Theorem \ref{theo:4.1}, it is easy
to build such generating matrix from the matrix $G_b$ for the Bose-Bush
code. For this case $G_d$ can be presented in the form
$G_d = [G\,|\,G^*(u_1,s_1)\,|\cdots|\,G^*(u_{h-2},s_{h-2})]$.
Here $G^*$ is obtained from $G_b$ by deleting of the first column, and
then $G^*(u_i,s_i)$ is obtained from $G^*$ by multiplying the second
row of $G^*$ by $\alpha^{u_i}$ and
the third row of $G^*$ by $\alpha^{s_i}$. The existence of such natural
numbers $u_i$ and $s_i$ follows from the existence of such codes
\cite{D69,CK86}.

These codes have IA: $\{(q-1)n,(q+1)(h-1)(q-h+1);1,(h-1)n\}$.
\end{enumerate}

\section{The main theorem}\label{main}

Now we can formulate and prove the main result of the paper.

\begin{theo}\label{theo:4.3}
Let $C$ be a linear nontrivial completely regular $[n,k,d]_q$-code
$C$ with covering radius $\rho = 2$, whose dual code $C^\perp$
is antipodal, i.e. a $[n,n-k,\{d^\perp,n\}]_q$-code $C^\perp$. Then
this code $C$ belongs to one of the families $(CR.1) - (CR.6)$ above.
\end{theo}

\pr
See \cite{BDZZ22} where additive $(n,N,\{d,n\})_q$-codes are classified and, for the linear case, correspond to the dual codes of the families $(CR.1) - (CR.6)$. In order to make the paper self-contained, we repeat here the arguments of \cite{BDZZ22} for the linear case.

Since $C$ is a nontrivial linear code it has cardinality
$N = q^k$, where $2 \leq k \leq n-2$.

We start from the case $k = 2$. For any prime power $q=p^s$ the
existence of $r$ mutually orthogonal Latin squares implies the
existence of a MDS $[r+2,2,r+1]_q$-code whose dual code
is a CR $[r+2,r,3]_q$-code from the family $(CR.3)$ (see Remark \ref{rem:2}).
These codes include the shortest nontrivial difference matrix
$[q,2,q-1]_q$-codes $(CR.2)$ which exist for any prime power $q$ and coincide
with Latin square codes.

Now consider the case $N = q^3$.
First show that in Denniston codes the number $h$ should divide $q$.
From Theorem \ref{theo:4.2} we have that $n$ is a multiple of $n-d$,
hence $n$ can be written as $n = (n-d)\ell$ for some natural number
$\ell$.
Hence $d=n(\ell-1)/\ell$, and we obtain from (\ref{dbound}) that
\[
 d = n \cdot \frac{\ell-1}{\ell} \leq n \cdot \frac{q-1}{q},
\]
which implies that $\ell \leq q$. But the case $\ell=q$ gives a difference matrix
code, we conclude, therefore, that $\ell < q$. Now assume that $n=1+(q+1)(h-1)$
and $d=q(h-1)$ for some natural number $h \geq 2$. It gives that $n=q(h-1)+h=d+h$.
So combining two equalities $n = 1+(q+1)(h-1)$ and $d = q(h-1) = n(\ell-1)/\ell$,
we obtain
\[
q(h-1) = (q(h-1) + h) \cdot \frac{\ell-1}{\ell},
\]
which implies that $h(\ell-1) = q(h-1)$. Since $h \geq 2$, and, hence,
$h$ and $h-1$ are mutually prime to each other, we deduce that $h$ divides
$q$, implying that we obtain a Denniston (or Bose-Bush for $h=2$) code.
%
%
%

The case $N = q^4$ gives also linear codes
from the family $(CR.4)$ \cite{SZZ69}. Indeed, having such a
$[n,4,\{d,n\}]_q$-code $C$ we can form the code $C_0$ which is a linear equidistant
$[n-1,3,d]_q$-code of length $n-1=(q^4-1)/(q-1)$ and $d = q^3$,
whose dual is a $q$-ary perfect Hamming code.

Now show that for this case $N = q^4$ there are no
codes of Bose-Bush, Denniston or Delsarte types. By Theorem \ref{theo:4.1}
the length of Bose-Bush or Denniston code should be $n = s (q^3-1)/(q-1) + 1$. Since
$n$ is a multiple of $n-d$ (see again Theorem \ref{theo:4.1}) the length
looks as $n = d \ell/(\ell-1)$ for some natural $\ell \leq q$. We obtain
(taking into account that $cd = s q^2$)
\EQ\label{posseq}
n = s \cdot\frac{q^3-1}{q-1} + 1 = d \cdot \frac{\ell}{\ell-1} = s q^2 \cdot \frac{\ell}{\ell-1}
\EN
Now we have to consider the case $(s,q-1) = 1$ and $(s,q-1) \geq 2$
separately.

Let, first, $(s,q-1) = 1$. Then we see, that  the left integer $n$ is not
divisible by $s$ and by $q$, but
the right integer is divisible by $s$ and $q$. We conclude that such
type codes do not exist for this case.

Now $(s,q-1) \geq 2$. For the case $s = q-1$ we obtain that $n = q^3$ and since
$N=q^4$, i.e. $N = q n$ we deduce that $C$ is a difference matrix code $(CR.2)$.

Now let $s=u(q-1)$ where $u \geq 2$. Using this $s$ in (\ref{posseq}) we arrive
to the expression
\[
0 \leq uq^2(q-\ell) = -u\ell+(u+\ell) - 1 \leq -1,
\]
which is impossible, since $2 \leq \ell \leq q$ and $u \geq 2$, which finishes the
case $N=q^4$.
%
%

The case $N=q^k$ for $k \geq 5$ can be considered in a similar way and we
omit these cases.
Now we have to say concerning the case when $q$ is a power of an
odd prime.

For the all codes $(CR.1)$ - $(CR.3)$ the proofs, which we gave,
work for even and odd case of prime power $q$ without any changes.

Bush in 1952  proved nonexistence of
$(q+2,3,q)_q$ codes \cite{B52} for odd $q$, i.e. $(CR.4)$-codes,
which implies the non-existence of Delsarte
$[q(q-1)/2, 3, \{q(q-2)/2, q(q-1)/2\}]_q$-codes $(CR.5)$-codes
for odd $q$ (since they are projective dual to Bose-Bush codes \cite{CK86}).
Then in 1997 Ball, Blokhuis and Mazzocca \cite{BBM97} proved the
non-existence of maximal arcs in Desarguesian planes of odd order,
which implies the non-existence of Denniston codes $(CR.6)$-codes
for odd $q$.
\qed

\section*{Acknowledgmements}
This work has been partially supported by the Spanish Ministerio de Ciencia e Innovación  under Grant PID2022-137924NB-I00
(AEI/FEDER UE).
The research of
the second and third authors of the paper was carried out at the
IITP RAS within the program of fundamental research on the topic
"Mathematical Foundations of the Theory of Error-Correcting Codes"
and was also supported by the National Science Foundation of
Bulgaria under project no. 20-51-18002.

\end{document}